\documentclass{article}

\usepackage{arxiv}

\usepackage[utf8]{inputenc} 
\usepackage[T1]{fontenc}    
\usepackage{hyperref}       
\usepackage{url}            
\usepackage{booktabs}       
\usepackage{amsfonts}       
\usepackage{nicefrac}       
\usepackage{microtype}      
\usepackage{lipsum}

\title{A pure connection formulation with real fields for Gravity}

\author{
  J. E. Rosales-Quintero \\
  Preparatoria 2 de octubre de 1968, Universidad Aut\'onoma de Puebla,\\ P.O. Box
1364, 72000 Puebla, México\\
  \texttt{ jeduardo.rosales@correo.buap.mx, jose.rosalesquintero@viep.com.mx} \\
}

\begin{document}
\maketitle

\begin{abstract}
We study an $SO(1,3)$ pure connection formulation in four dimensions for real-valued fields, inspired by the Capovilla, Dell and Jacobson complex self-dual approach. By considering the CMPR BF action, also, taking into account a more general class of the  Cartan-Killing form for the Lie algebra $\mathfrak{so(1,3)}$  and by refining the structure of the Lagrange multipliers, we integrate out the metric variables in order to obtain the pure connection action. Once we have obtained this action, we impose  certain restrictions on the Lagrange multipliers, in such a way that the equations of motion led us  to a family of torsionless conformally flat Einstein manifolds,   parametrized by two numbers. Finally, we show that, by a suitable choice of parameters, that  self-dual spaces (Anti-) De Sitter can be obtained.
\end{abstract}

\keywords{Gravity \and Pure Connection Actions  \and Einstein manifolds \and Self-dual spaces  }

\section{Introduction}
Spacetime can be described as a four dimensional pseudo-Riemannian Lorentzian manifold with curvature described  by the metric. As such, the metric is the fundamental object used to describe gravity, it equips the causal structure and ascertains the spacetime intervals. This metric formulation can be constructed from a variational principle,  known in the literature as the second order formulation. It is also possible to construct a first order formulation where we consider the  connection  as independent field along with the metric field itself, the well-known first-order formulation. The connection is introduced as an auxiliary object which takes the second-order in derivatives for the original action to a formulation which consist in derivatives of first order. The appearance of the connection not only permits a simplification on the differential equation, it also has a deep meaning, it let us construct principal fibre bundles over the spacetime, then we can consider gravity  as a gauge theory where we can take different gauge groups   in $GL(4)$.\\
Given this context, it is surprising that we can reformulate gravity in such a way that the metric variable disappears completely  by  ''integrating it out'' and then, leaving the most important role as a fundamental field to the connection, then, as we do not need a pre-existing metric, the theory becomes background free. These kind of formulations are known in the literature as pure connection actions for gravity \cite{Peldan1990} (for a recent review see \cite{Krasnov2018}).\\
In the late 80's of the last century by Capovilla, Dell and Jacobson\cite{Capovilla1990}\cite{Capovilla1991}\cite{Capovilla1991a} built a pure connection formulation for gravity (CDJ Action) by taking as the fundamental field a complex $su(2)$ connection. However, the Lagrangian, contains an additional auxiliary field of density minus one. Then it was realized that in this formulation, General Relativity (GR) is not unique, this means that  there is not a single diffeomorphism invariant gauge theory that shares the
same key properties with GR such as  they have two propagating polarizations of the graviton.
This was found by Capovilla et. al.  through a two parameter family of ''neighbors'', and then,  in the same spirit, by an infinite parameter of theories, found by  Bengtsson\cite{Bengtsson1991} and that later on, rediscovery by Krasnov\cite{KRASNOV2007}.\\
In this work, we considered an $SO(3, 1)$ modified CDJ pure connection action for four dimensional gravity, inspired  by the paper by Rosales-Quintero\cite{Rosales-Quintero2016}, there, it's considered an $SU(2)$ gauge invariant pure connection action where it's imposed extra algebraic constraints in order to let it suitable  for a straightforward $OSp(1,2)$ supersymmetric (SUSY) generalization.  These constraints put some  algebraic relations over the dynamical fields, and cojointly with  integrability conditions and Bianchi identity, it's found that the only solutions are gravitational instantons, e. i., Einstein metrics with half of the Weyl curvature vanishing, and for the SUSY case, a corresponding antiself-dual supergravity $N=1$ where a spin $3/2$ field appeared naturally, and can be related to the supersymmetric partner of the graviton, the gravitino field. We have to mention that not all those extra constraints, considered in \cite{Rosales-Quintero2016} played a very important role in the pure bosonic case, and since in this works we will not consider the SUSY case, then we have to choose which constraints are the most important constraints to keep.\\
In order to identify the constraints that will be needed in this paper, we will start  from the formalism for $BF$ action for $SO(3,1)$ gauge group proposed by Capovilla et. al. (CMPR action)\cite{Capovilla2001} but before we integrate out the metric variable in this action,  we will give some details over the Cartan-Killing form and the auxiliary fields, so that  we'll obtain an adequate pure connection formulation from where maximally symmetric spaces will appear as solutions from its variational principles.\\  
The organization of the paper is as follows. In section 2, we present a brief introduction to $BF$ theory together with CMPR formulation for $SO(3, 1)$-valued fields. The formulation is presented in such a way that it let us construct our pure connection action. In section 3, it's presented the pure connection formulation and from it, we calculate the equations of motion, obtaining as solutions Einstein  spaces, i. e., spaces whose Ricci tensor is proportional to the metric, being this constant of proportionality proportional to the cosmological constant. These kind of spaces are also known in the literature as gravitational instantons. We conclude with a brief discussion and outlooks.\\
In this paper, we have labeled $so(1,3)$ Lie algebra indices by the
beginning of the Latin alphabet lowercase letters $\{a, b, c, \ldots \}$ and Greek alphabet letters for spacetime indices $\{ \mu, \nu, \rho, \ldots \}$. Also, we consider the Minkowski metric as $\eta_{ab}=diag(-1,1,1,1)$, and we establish $\eta_{ab,cd}=\eta_{ac}\eta_{bd}-\eta_{ad}\eta_{bc}$. Additionally, we define $G_{(ab)} = G_{ab} + G_{ba}$ and $G_{[ab]} = G_{ab} - G_{ba}$.

\section{From BF to pure connection formulation}
BF theory is topological action, which means that from this action, no degrees of freedom can be obtained and represents a theory which everywhere looks locally the same, i.e. a flat spacetime. 
It's  a gauge invariant action
in which its basic ingredients are a principal fiber bundle $P$
over the four dimensional orientable spacetime manifold $\mathcal{M}$ - without a metric structure  on it - together with a Lie group $G$
which acts as internal group, and whose Lie algebra $\mathfrak{g}$
is equipped with a non degenerate bilinear invariant form
$\kappa$, the Cartan-Killing form (CKf), and a $\mathfrak{g}$-valued
connection $A$ which defines the field strength $F$, and a two-form $\mathfrak{g}$-valued field $B$\cite{Baez2000}\footnote{In reality $A$ is a connection on P, and $B$ is an $ad(P)$-valued two-form. The definition given is precise if we choose a local trivialization. }\\
The usual way to get a theory for GR from this kind of action is by breaking explicitly its topological behavior by means of the introduction of some Lagrange multipliers which constraints the functional form of the two-form field B, the so-called simplicity constraints. In general, the action can be written as follows
\begin{equation}
    S[A, B, \Phi]=\int_{\mathcal{M}} tr(B\wedge F + \mathfrak{s}(B \wedge B))
\end{equation}
where ''$tr$'' is the CKf pairing on $g$, and $\mathfrak{s}(B\wedge B)$ encodes the simplicity constraints. At the middle of the seventies of the last century, Pleba\'nski considered self-dual quantities by a  complexified $SO(3)$ group, but in order to describe Lorentzian signature spacetime it was necessary to impose certain reality conditions\cite{Plebanski1977}. Here, we consider one of the most well-known BF action functional which give rise to an action for GR with Immirzi parameter\cite{Capovilla2001}, by considering $SO(1,3)$ as our internal symmetry group 
\begin{equation}  \label{eq: Action CMPR}
    S[A, B, \Phi, \mu]=\int_{\mathcal{M}} tr(B\wedge F -\frac{1}{2}\ \Phi(B)\wedge B+\rho (a_{1}\Phi+2a_{2}*\Phi)).
\end{equation}
where $a_{1}$ and $a_{2}$ are constants factors that can not vanish simultaneously. The fundamental fields are characterized by  $B=B^{ab}t_{ab}$ and $A=A^{ab}t_{ab}$, so the field strength is given, as usual by, $F=dA+A\wedge
A=(dA^{ab}+A^{ac}\wedge A_{c}^{\ b})t_{ab}$, being $t_{ab}$ the $\mathfrak{so(1,3)}$ Lie algebra generators, satisfying
\begin{displaymath}
  [t_{ab},t_{cd}]=-\frac{1}{4}\ \eta_{ab,[c}^{\ \ \ \ \ [e}\eta_{d]}^{\ \ f]} t_{ef}
\end{displaymath}
so the CKf is calculated straightforward as
\begin{equation}
    \kappa_{ab,cd}=\frac{1}{I_{ad}}\ tr(t_{ab}t_{cd})=\frac{1}{2}\ \eta_{ab,cd}.
\end{equation}
In addition,  $\Phi$ is a Lagrange multiplier which behaves as a endomorphism  when  acts over the B field,  $\Phi(B)=\Phi^{abcd}B_{ab}t_{cd}$,  otherwise it is a Lie algebra bivector valued field $\Phi=\Phi^{abcd}t_{ab}t_{cd}$ , $\rho$ is also a Langrange multiplier that constraint the functional form of the B-field and $*$ is the Hodge dual acting on a pair of $\mathfrak{so(1,3)}$ Lie algebra valued fields, defined as $*\xi^{ab}=\frac{1}{2}\ \epsilon^{abcd}\xi_{cd}$. Then action (\ref{eq: Action CMPR}) can be rewritten as
\begin{equation} \label{eq: Action CMPR developed}
    S[A. B, \Phi, \mu]=\int_{\mathcal{M}} B^{ab}\wedge F_{ab} -\frac{1}{2}\  \Phi^{abcd}B_{ab}\wedge B_{cd}+\rho (a_{1}\Phi^{ab}_{\ \ \ ab}+a_{2}\epsilon_{abcd}\Phi^{abcd}). 
\end{equation}
The simplicity constraints are given by the equations of motion coming from the Lagrange multipliers $\Phi$ y $\rho$, and they imply that $B=\alpha e\wedge e + \beta *(e\wedge e)$ where $e$ is the one-form tetrad field, $\alpha$ and $\beta$ are real parameters that satisfy $\alpha^2 \neq \beta^2$ and the ratio $\alpha/\beta$ is determined by the ratio
\begin{equation} \label{eq: Ratio a2a1}
a_{2}/a_{1}=(\beta^2-\alpha^2)/4\alpha\beta
\end{equation}
Finally if we substitute back the functional form of the $B$ field into the action (\ref{eq: Action CMPR developed}), we obtain the Holst action\cite{Holst1996}
\begin{equation} \label{eq: Holst action}
    S[A, e]=\int_{\mathcal{M}} \alpha e^{a}\wedge e^{b}\wedge F_{ab}+  \beta \epsilon_{abcd} e^{a}\wedge e^{b}\wedge F^{cd}.
\end{equation}
where we can identify the Immirzi parameter, $1/\gamma=\alpha/\beta$. Now, let us construct the pure connection action, by considering the equation of motion coming from the variation on $B$ in (\ref{eq: Action CMPR developed}):
\begin{equation}
    \delta_{B}S=0 \qquad \Rightarrow \qquad F_{ab}=\Phi_{abcd} B^{cd}
\end{equation}
and if we also consider that det$\Phi\neq 0$ then $B_{ab}=(\Phi^{-1})_{abcd}F^{cd}$. Let us substitute back this result in (\ref{eq: Action CMPR developed})
\begin{equation} 
    S[A, \Psi, \Phi, \rho]=\int_{\mathcal{M}} tr\bigg(\frac{1}{2}\ \Psi(F)\wedge F+\rho (a_{1}\Phi+2a_{2}*\Phi)\bigg)
\end{equation}
where we have defined $\Psi=\Phi^{(-1)}$ which has the same behavior as $\Phi$, it is an endomorphism acting on Lie algebra valued two-forms or an algebra bivector field. Further, following \cite{Rosales-Quintero2016}, we observe that the terms related to the simplicity constraints, are  algebraic-type, and its work is to constraint the functional form of $F$, then we have to rewrite the restrictions over $\Psi$ in such a way they maintain the same algebraic properties. But before we present the final form of the pure connection action, it will be  necessary to take into account some details over some of the elements that we have used in the action (\ref{eq: Action CMPR}):
\begin{itemize}
    \item In $\mathfrak{so(1, 3)}$ there are two invariant traces that can be
defined, so instead of just consider  the usual trace of the Lie algebra generators in the adjoint representation, what we have defined as $tr$, we will consider the most general case\cite{Alexandrov2007}, so we
define the trace between two algebra generators  as
\begin{displaymath}
Tr(t_{ab}t_{cd})=a_{1} \kappa_{ab,cd}+2a_{2}\ *\kappa_{ab,cd}
\end{displaymath}
where $a_{1}$ and $a_{2}$ are the same constants considered in the action (\ref{eq: Action CMPR}). The last equation gives
\begin{equation} \label{eq: General trace definition}
Tr(t_{ab}t_{cd})=\frac{1}{2}\ \bigg( a_{1}\eta_{ab,cd}+a_{2}
\epsilon_{abcd} \bigg):=\Pi_{ab,cd},
\end{equation}
where have renamed $Tr(\ )$ as $\Pi$, in order to associated it to a
pseudoprojector of $\mathfrak{so(1, 3)}$ (the
projector property is achieved if only if $a_{1}=1/2$ and $a_{2}=\mp i/2$, the self-dual and anti-self-dual sectors respectively), and
its inverse is calculated by the relation $(\Pi\Pi^{-1})_{ab}^{\ \
cd}=(\Pi^{-1}\Pi)_{ab}^{\ \ cd}=\frac{1}{2}\ \eta_{ab}^{\ \ cd}$,
implying that
\begin{equation}  \label{eq: Inverse pseudoprojector}
\Pi^{-1}_{abcd}=\frac{1}{2}\ \frac{1}{a_{1}^{2}+a_{2}^{2}}\
(a_{1}\eta_{ab,cd}-a_{2}\epsilon_{abcd})
\end{equation}
where $a_{1}^{2}+a_{2}^{2}\neq 0$.

    \item Let us define the endomorphism $\Phi(B)$ as follows:
    \begin{equation}
        \Phi(B)=\Phi^{abcd} B^{ef}\cdot tr(t_{ab}t_{ef})t_{cd}=\Phi^{abcd} B_{ab}t_{cd}
    \end{equation}
    but now, instead of taking the usual trace "tr", let us use the trace defined in (\ref{eq: General trace definition}), then we obtain
    \begin{equation} \label{eq: Automorphism general action}
        \Phi(B)=\Phi^{abcd} B^{ef}\cdot Tr(t_{ab}t_{ef})t_{cd}=\Phi^{abcd} B^{ef} \Pi_{abef}t_{cd}.
    \end{equation}
    We will apply the same results for the $\Psi$ field since we have considered that it inherit the same algebraic and endomorphism behavior form of $\Phi$.
\end{itemize}
It is not difficult to show that the action (\ref{eq: Action CMPR}), together with the two arguments previously described,  maintains the same results, i.e., we still get the Holst action without some additional structure for the new changes. Finally, the pure connection action takes the final form
\begin{equation} \label{eq: Rosales Action}
    S[A, \Psi, \rho]=\int_{\mathcal{M}} Tr(\Psi(F)\wedge F+\rho\Psi).
\end{equation}
Furthermore, we  note that the last action, inherits similar diffeomorphism and  gauge symmetry
as in the $BF$ theory. It also has identical topological spirit, in the sense that it does not
involve background structure and there are not metric variables of any kind. This is the action that we will consider for the forthcoming  rest of the paper, forgetting that it comes from the "BF" action and taking it as an independent one. 


\section{ Pure Connection Action for Gravity}
Now, it is time to obtain the equations of motion for the action (\ref{eq: Rosales Action} ). But first, let us consider (\ref{eq: General trace definition} ) and (\ref{eq: Automorphism general action}), hence the actions reads
\begin{equation}   
S_{FF}[A, \Psi, \rho ]=\int_{\mathcal{M}} 
 \Psi^{abcd}\mathfrak{F}_{ab} \wedge \mathfrak{F}_{cd} + \rho \Psi^{abcd}\Pi_{ab,cd}
\end{equation}
where we have defined $\mathfrak{F}^{ab}=\Pi^{abcd}F_{cd}$.  The equation
of motion coming from $\rho$ allow us to decompose the $\Psi$ field into its irreducible components, the trace and traceless part, respectively, as follows
\begin{displaymath}
  \Psi^{abcd}=
  \bigg[ \frac{1}{6}\ \eta_{ab,cd} \Psi^{ef}_{\ \ ef}-\frac{1}{24}\ \epsilon^{abcd}\Psi^{efgh}\epsilon_{efgh}    \bigg] + \bigg[ \Psi^{abcd}-\frac{1}{6}\ \eta_{ab,cd} \Psi^{ef}_{\ \ ef}+\frac{1}{24}\ \epsilon^{abcd}\Psi^{efgh}\epsilon_{efgh}   \bigg]
\end{displaymath}
\begin{equation}
 =\frac{\Psi^{ef}_{\ \ ef}}{12 a_{2}}(2 a_{2}\eta^{ab,cd}+a_{1}\epsilon^{abcd})+\Psi^{T \ abcd}.
\end{equation}
The action is rewritten as
\begin{equation}
  S_{FF}=\int_{\mathcal{M}} \Psi^{T \ abcd} \mathfrak{F}_{ab} \wedge \mathfrak{F}_{cd}
  +\frac{\Psi^{ef}_{\ \ ef}}{12 a_{2}}(2 a_{2}\eta^{ab,cd}+a_{1}\epsilon^{abcd}) \mathfrak{F}_{ab} \wedge \mathfrak{F}_{cd}.
\end{equation}
The equations of motion coming from the variation of the $\Psi$ field are given by
\begin{eqnarray}
\label{eq: CMPR constraint FF1}   
\mathfrak{F}^{ab} \wedge \mathfrak{F}^{cd} & = & \frac{1}{6}\  \eta^{ab,cd}\mathfrak{F}^{fg}\wedge \mathfrak{F}_{fg}-\frac{1}{24}\ \epsilon^{abcd} \mathfrak{F}^{ef}\wedge \mathfrak{F}^{gh}\epsilon_{efgh} \\
\label{eq: CMPR constraint FF2}            4 a_{2}\mathfrak{F}^{ab}\wedge \mathfrak{F}_{ab}   & = &-a_{1}\epsilon^{abcd} \mathfrak{F}_{ab}\wedge \mathfrak{F}_{cd}, 
\end{eqnarray}
the last equations are the simplicity constraints, which let us write the two-form field $\mathfrak{F}$ as the product of two one-forms as follows
\begin{equation}   \label{eq: simplicity}
\mathfrak{F}^{ab}=\Lambda(\alpha \Sigma^{ab}+\beta *\Sigma^{ab}) \ \ \ \textrm{where}\ \ \ \Sigma^{ab}=e^{a}\wedge e^{b}.
\end{equation}
where $\Lambda$ is the cosmological constant, and $\alpha$, $\beta$ are constant dimensionless terms  satisfying the relation (\ref{eq: Ratio a2a1}). We have to point out that in the original BF-formulation the cosmological constant term does not enter into play along with the solution of the simplicity constraints but here, instead, it's introduced  due to dimensional consistency\cite{Capovilla1990}.\\
We observe that if we apply the covariant derivative on both sides in (\ref{eq: simplicity}), and by using the Bianchi identity, we have
\begin{equation} \label{eq: zero torsion condition}
  D(\alpha\Sigma^{ab}+\beta*\Sigma^{ab})=0  \ \ \Rightarrow \ \ T^{a}=De^{a}=0,
\end{equation}
where $D$ is the covariant derivative, defined as usual as $D=d+[A,\ ]$. Equation (\ref{eq: zero torsion condition}) is the well-known, zero torsion condition, which let us write the connection as function of the tetrad field, $A=A(e)$.\\
But before going through the equation of motion for the connection, we have to observe an important point, if we consider the auto-self dual formulation of $SO(3)$ instead of the Lorentz group,  and instead of the pseudo projector $\Pi$ we take the anti-self-dual projector, then   (\ref{eq: simplicity}) is the instanton solution,i.e., solutions of the Einstein's field equations together with a vanishing self-dual Weyl tensor\cite{Krasnov2017}. In our case, the instanton solution-type  can be seen clearly if we apply the inverse pseudo-projector (\ref{eq: Inverse pseudoprojector}) to (\ref{eq: simplicity}), which give us the equation
\begin{equation}   \label{eq: F full}
  F_{ab}=\frac{1}{2(a_{1}^2+a_{2}^2)}\bigg[ (a_{1}\alpha+a_{2}\beta)\eta_{ab}^{\ \ cd}+(a_{1}\beta-a_{2}\alpha)\epsilon_{ab}^{\ \ cd}    \bigg]\Sigma_{cd}
\end{equation}
but by applying the covariant derivative to  the zero torsion condition,  $D T_{a}=F_{ab}\wedge e^{b}=0$,  (\ref{eq: F full}) means
\begin{equation}   \label{eq: Fundamental relation}
    a_{1}\beta-a_{2}\alpha=0 \qquad \Rightarrow \qquad \frac{a_{1}}{a_{2}}=\frac{\alpha}{\beta},
\end{equation}
 since  we are not considering degenerate tetrad fields. The last relation together with (\ref{eq: Ratio a2a1}) indicates 
\begin{equation}
  \beta^2=-\frac{1}{3}\ \alpha^2   \qquad  \textrm{and}    \qquad a_{2}^2=-\frac{1}{3} a_{1}^{2}
\end{equation}
and the $\mathfrak{so(1,3)}$ valued field strength is written as function of only two free parameters $\{ \alpha, a_{1} \} $ or $\{ \beta, a_{2} \} $, as follows
\begin{equation} \label{eq: Einstein Manifolds}
 F_{ab}=\frac{\alpha}{a_{1}}\ \Lambda\Sigma_{ab}=\frac{\beta}{a_{2}}\ \Lambda\Sigma_{ab}.
\end{equation}
Now, without loss of generality, it is possible to consider one of the   quotients $\alpha/a_{1}$ or $\beta/a_{2}$ as real since $a_{1}, a_{2}, \alpha, \beta$ come from algebraic type equations and they're  defined upon an overall constant factor. Then if we choose one quotient real, then  relation (\ref{eq: Fundamental relation}) implies directly that the another one is real too. Equation (\ref{eq: Einstein Manifolds}) can be rewritten as $Ric=\lambda g$ where $Ric$ is the Ricci tensor built by tracing the field tensor with the metric tensor  $g_{\mu\nu}=e_{\mu}^{\ a}e_{\nu a}$.  Such Pseudo-Riemannian manifolds are known in the literature as Einstein manifolds and since the Weyl tensor vanishes then the solutions of (\ref{eq: Einstein Manifolds}) are conformally flat spaces\cite{Besse1987}, also these manifolds are maximally symmetric spaces, i.e., they are  homogeneous and isotropic spaces. But even more, from (\ref{eq: Einstein Manifolds}), if we consider $\alpha/a_{1}=1/3$ then 
\begin{equation}
    F_{ab}=\frac{\Lambda}{3}\ e_{a}\wedge e_{b}
\end{equation}
we can identify Anti-De Sitter or De Sitter spaces together with flat spaces (they are trivially obtained if $\alpha$ and $\beta$ vanish simultaneously) depending on the sign of the cosmological constant. When we have nonvanishing $\Lambda$ this equation defines self-dual solutions\cite{Alexander2019} and coincides with the descriptions given for Einstein manifolds\cite{Cotsakis-Gibbons}.\\  
Finally,  the equation motion coming from the variation of the connection $A$. It is given by
\begin{equation}    \label{eq: Integrability condition}
  D\Psi^{abcd}F_{cd}=0
\end{equation}
which implies that $\Psi^{abcd}F_{cd}$ is covariantly constant and it's a differential equation that gives rise, if we apply  a second covariant derivative,  to an algebraic equation called integrability condition\cite{Rosales-Quintero2016},  this condition constraints the behavior of the $\Psi$ field. This constraint doesn't give  any significance result, at least at pure bosonic case, but if we want to consider a supersymmetric extension case, then the integrability condition becomes a very important tool. The supersymmetric extension goes beyond the scope of this paper and it will be presented in a further work.

\section{Conlusions and Outlooks}
As was noted, by Capovilla et. al.,  at the ending of the 80's of the last century, and a few years later by Bengtsson, there is a family of diffeomorphism invariant pure connection gauge theories that share the same key properties with GR. Among  the elements in this family, we considered in this work, a pure connection action inspired in the CDJ formulation and considering the modification proposed by Rosales-Quintero\cite{Rosales-Quintero2016}. In the last reference, the action its constructed in such a way, that its equation of motions imply (anti) self-dual Einstein spaces, where, also the Weyl tensor vanishes, at least its (anti)self-dual components. Moreover, in \cite{Rosales-Quintero2016} it is considered more algebraic constraints, since the mean idea was to build up a supersymmetric extension of the theory. In this work, we were considering the pure bosonic case then we took into consideration only   the constraint over the trace terms of the $\Psi$ field. Then,  we also, had to refine the definition of some of the components used for the construction of the action, such as the CKf and the endomorphism $\Psi(F)$. Once we have done this, we observed, that, from the equation of motion of the $\Psi$ field, that we obtained Einstein spaces, in fact, a family of them, parametrized by quotient $\alpha/a_{1}$ or $\beta/a_{2}$,   that are conformally flat. But even more, if we choose the quotients to be $1/3$ then we obtain, self-dual solutions\cite{Alexander2019}.\\
Since we have investigated just the pure bosonic case, it will be interesting to consider also a supersymmetric extension as was proposed originally in (ref a mi), for $N=1$. Also, we have not consider the relation with the Immirzi parameter, at least explicitly, and it could be interesting to consider in a future work.\\
It will be interesting to check the relation between  (\ref{eq: Einstein Manifolds}) and the quasitopological principle proposed by Alexander et. al.\cite{Alexander2019} and their $\theta$ term for the non-constant cosmological constant $\Lambda$.\\ 
Considering the canonical analysis of the action for the ASD complex case and in the $\mathfrak{so(1, 3)}$ formulation, will be important for a canonical quantization programme. \\


\section{Acknowledgments}
The author acknowledges support from Benem\'erita Universidad Aut\'onoma de Puebla  for granting the necessary conditions to carry out this research work.


\bibliographystyle{unsrt}  
\bibliography{references}  


%



\end{document}